\pdfoutput=1
\documentclass[conference]{IEEEtran}
\usepackage{blindtext, graphicx}

\ifCLASSOPTIONcompsoc
  \usepackage[caption=false,font=normalsize,labelfont=sf,textfont=sf]{subfig}
\else
  \usepackage[caption=false,font=footnotesize]{subfig}
\fi

\usepackage[utf8]{inputenc}
\usepackage[english]{babel}
\usepackage{array}
\usepackage{tabu}
\usepackage{longtable}
    \usepackage{booktabs,tabularx}

\usepackage{wrapfig}
\usepackage{graphicx}
\usepackage{parskip}
\usepackage{gensymb}
\usepackage{sidecap}
\usepackage{amsmath}
\usepackage{MnSymbol}
\usepackage{wasysym}
\usepackage{avm}

\graphicspath{ {images/pdf/} }

\usepackage{tikz}
\usepackage{textcomp}
\usepackage{hyperref}
\usepackage{lipsum}

\newcommand\copyrighttext{%
  \footnotesize \textcopyright 2017 IEEE. Personal use of this material is permitted.
  Permission from IEEE must be obtained for all other uses, in any current or future
  media, including reprinting/republishing this material for advertising or promotional
  purposes, creating new collective works, for resale or redistribution to servers or
  lists, or reuse of any copyrighted component of this work in other works.
  }
\newcommand\copyrightnotice{%
\begin{tikzpicture}[remember picture,overlay]
\node[anchor=south,yshift=10pt] at (current page.south) {\fbox{\parbox{\dimexpr\textwidth-\fboxsep-\fboxrule\relax}{\copyrighttext}}};
\end{tikzpicture}%
}

\begin{document}
\title{A Compact, Wide Field-of-View Gradient-index Lens Antenna for Millimeter-wave MIMO on Mobile Devices}

\author{\IEEEauthorblockN{Wenlong Bai and Jonathan Chisum}
\IEEEauthorblockA{
Electrical Engineering Department\\
University of Notre Dame\\
Notre Dame, IN 46556\\
Email: wbai2@nd.edu, jchisum@nd.edu}
}

\maketitle
\copyrightnotice

\begin{abstract}
Lens-based beam-forming antennas offer a low-power, low-cost alternative to hybrid beamforming antenna arrays. They are ideally suited to millimeter-wave massive MIMO systems due to their native beam-space operation and angular selectivity and minimal dependence of high-speed data converters. We discuss the design of a compact and low-cost lens-based beam-forming antenna for small form-factor platforms such as small-cells and mobile devices in 5G wireless networks. We discuss a gradient-index design method and low-cost fabrication method based on perforated dielectrics. We discuss the need for high-contrast permittivity ranges to achieve wide scan angles which are essential for leveraging the full capability of massive MIMO systems (e.g., full stream capacity). Finally, we show that by using an appropriately designed perforated medium, gradient-index lenses with low minimum permittivity of $1.25$ can achieve a maximum beam-steering angle of $44$ degrees. We suggest that such an approach can enable practical low-loss, low-cost, and compact beam-steering lens antennas for millimeter-wave MIMO with wide beam-steering angles.
\end{abstract}

\IEEEpeerreviewmaketitle


\section{Introduction}

\IEEEPARstart
With the recently proposed FCC allocations for mobile wireless in the millimeter-wave (mmWave) bands, fifth generation (5G) wireless communications now seems inevitable. The move to mmWave brings with it the promise of wideband channels and almost unlimited spatial reuse \cite{hur_krogmeier_lov_ghosh_mmWavebeamforming}. However, to take advantage of this underutilized spectrum several significant challenges must be overcome. The dominant characteristic of millimeter-wave propagation is increased path loss. To realize a practical network under such constraints antenna arrays with high gain, beam-steering, and massive MIMO processing have been proposed \cite{hur_krogmeier_lov_ghosh_mmWavebeamforming,rappaport_itwillwork}.

Traditional beam-steering approaches include digital base-band beamforming, analog RF/LO beamforming, and hybrid beamforming. While full digital beamforming offers the most flexibility it requires a data-converter and RF transceiver on every antenna element and is therefore prohibitive at 5G channel bandwidths\cite{zeng_mmWaveMIMO_TAC}. Analog beamforming requires only a single data-converter and transceiver but only has one stream. Hybrid beamforming \cite{Alkhateeb_heath_HybridBF} has emerged as a compromise between digital and analog beamforming--in an $N-$element antenna array, $M<<N$ baseband data-converters can be combined with $N$ RF phase shifters to provide high performance beam-steering, high-gain, and $M$ independent data streams. However, even hybrid beamforming has its limitations, namely cost, power and sensitivity to analog impairments due to finite-resolution, lossy mmWave phase shifters. In addition, signals from multiple angles of arrival (AoA) are incident upon all receive chains and so linearity requirements are increased.

One of the main motivations for moving to a large beam-steering array is to leverage the benefits of massive MIMO including high-gain and spatial multiplexing \cite{hogan_sayeed_MIMO,larsson_marzetta_MassiveMIMO}. Both are native beam-space concepts and so lens antennas, fundamentally beam-space devices, have been proposed as ideal apertures for mmWave MIMO \cite{zeng_mmWaveMIMO_TAC}. The key advantage of lenses is inherent angle-, or beam-space selectivity. With a single RF chain a stream incident from any angle can be received with high gain. Reception of a specific beam from an AoA requires minimal processing and is realized in the completely passive lens medium. With $M$ RF chains, $M$ independent streams from $M$ angles of arrival can be received. Due to significant attenuation of multipath components, the typical number of paths (and therefore approximately the number of AoAs) in mmWave bands is $L=2-8$ \cite{rappaport_itwillwork} therefore a lens antenna with 2-8 transceivers and data-converters can capture all significant paths from any AoA up to the field of view (FOV) of the antenna. In \cite{zeng_mmWaveMIMO_TAC} it was noted that an ideal lens antenna can achieve maximum MIMO capacity equal to $L$ times the channel capacity where $L$ is the number of spatial-multiplexed streams. A practical lens can approach this value only if every spatial stream emitted by the transmitter is intercepted by the receiver. Therefore, it is clear than any lens antenna for mmWave MIMO applications should have a large FOV to capture all $L$ streams. 

In this paper we present a beam-forming lens antenna with wide FOV for use in small-form-factor platforms such as small-cells and even mobile devices. The lens is designed with transformation optics \cite{kwon_transformationEM_2010} and realized as a gradient index (GRIN) lens using stacked layers of perforated dielectric as we recently demonstrated in \cite{garcia_apsursi}. Because our process is based on photolithography, etching is parallel and therefore arbitrarily complex lenses can be fabricated for no additional cost and with negligible effect on fabrication time. From start to finish, each layer (wafer) of the GRIN lenses requires between 1-2 hours of etch time and since wafers are etched in batches, an entire lens can be etched in the same amount of time.

\section{Wide Field-of-view Lens Concept}

\begin{figure*}[!ht]
    \centering
    \includegraphics[height=0.425\textwidth]{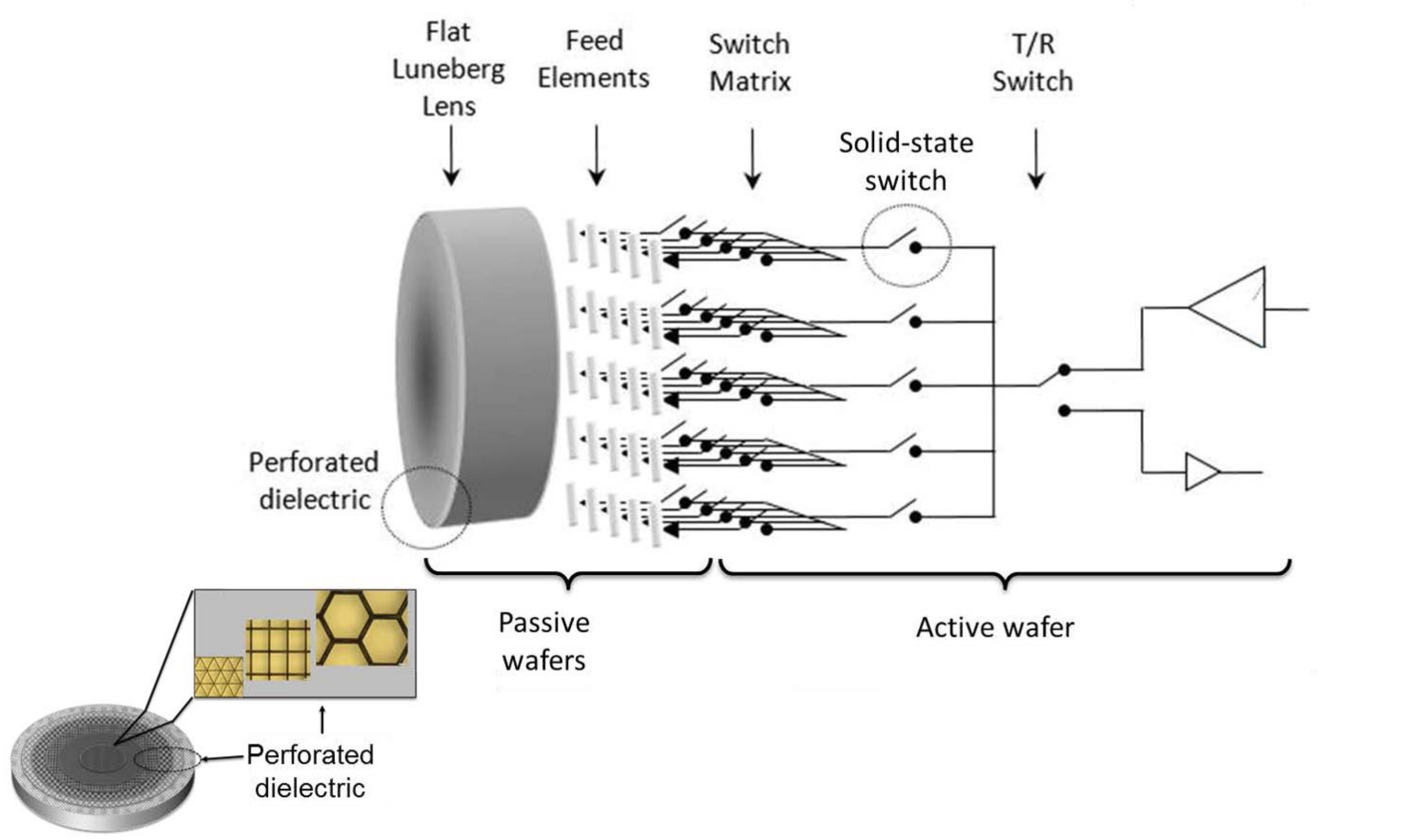}
    \caption{A flat lens antenna with switch-beam feed network provides a low-complexity millimeter-wave beam-steering antenna for inclusion in low-power, low-cost small cells and mobile devices.}
    \label{fig:antennaconcept}
\end{figure*}

Figure\,\ref{fig:antennaconcept} shows the lens antenna concept. The lens design is based on a flat Luneburg lens (FLL) fabricated with a stack of perforated silicon wafers. The FLL provides beam-steering by directing the RF signal to a particular feed element. The feed element ($x,y$) location dictates the beams ($\theta,\phi$). The feed network is driven by an RF stream (single stream shown for clarity) through a matrix of solid-state switches. Importantly, mmWave switches are simpler to realize, lower loss, and require less power than variable phase shifters. Note that the switch matrix is only shown for concept clarity and is not the focus of this paper. 

The FLL is designed with transformation optics \cite{kwon_transformationEM_2010}. Transformation optics (TO) is means of mapping a spatial distortion into a transformation of the material constitutive parameters (permittivity, $\epsilon_r$ and permeability, $\mu_r$). While a discussion of TO is beyond the scope of this report we summarize the approach: i) a coordinate transformation from physical space, ($x,y,z$) to virtual space ($x',y',z'$) is defined, ii) the transformation is used to modify the permittivity and permeability tensors of an original design, and iii) the result is a spatial map of $\epsilon_r$ and $\mu_r$ which realize the functionality of the original structure in the physically transformed device. 

\begin{figure}
    \centering
    \subfloat[Config. 1]{
        \label{fig:erorig}
        \includegraphics[width=0.5\columnwidth]{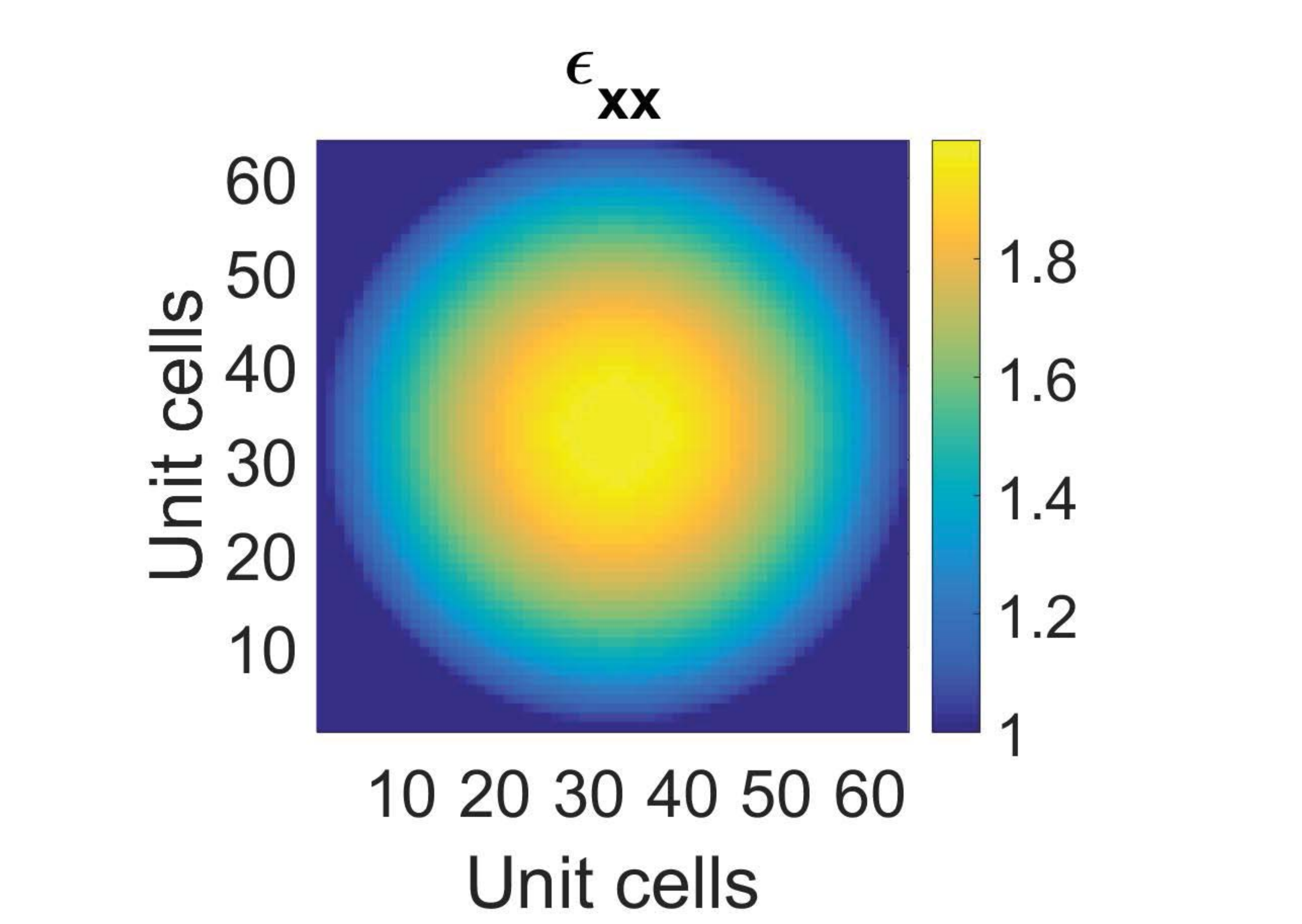}
        }
    \subfloat[Config. 2]{
        \label{fig:flatlens}
        \includegraphics[width=0.5\columnwidth]{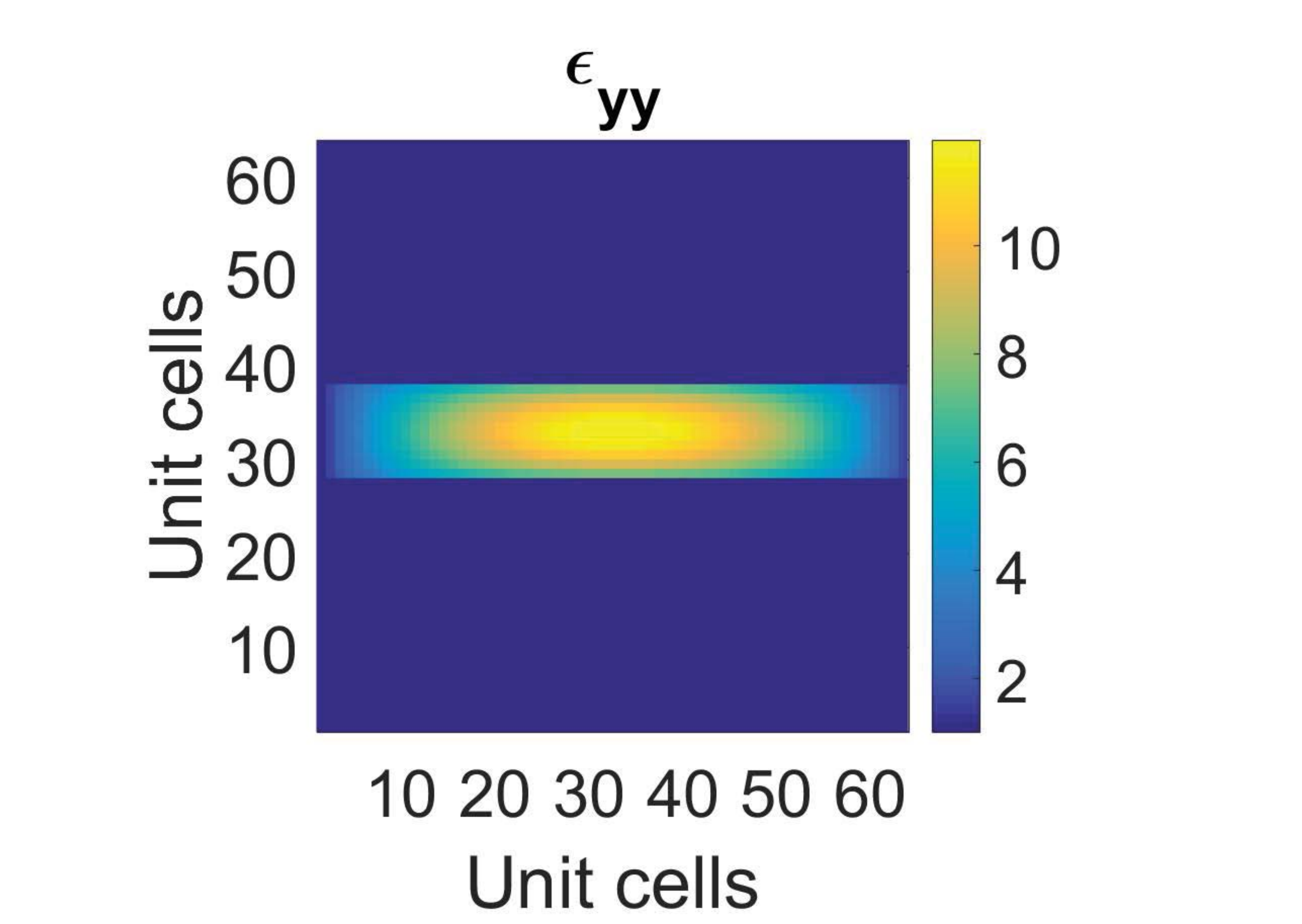}
        }
    \caption{(a) Permittivity map for the original Luneburg lens. (b) Permittivity map for the flat Luneburg lens. Both lenses have a radius of $23.1\,$mm corresponding to three wavelengths at the design frequency of $39\,$GHz and each unit cell is $0.72\textrm{mm} \times 0.72\textrm{mm}$.}
\end{figure}

We now apply this process to the Luneburg lens which is a spherical lens whose permittivity varies from $\epsilon_r=2$ at the center of the lens to $\epsilon_r=1$ at the surface as shown in Fig.\,\ref{fig:erorig}.
\begin{equation}
    \epsilon_r=\left(2-\frac{x^2+y^2}{R^2}\right)\cdot\textrm{eye}(3,3) \label{eq1}
\end{equation}

The spherical lens is converted to a cylinder with a compression ratio, $\delta$, which is introduced to compress the vertical, $y$, axis: 
\begin{align} 
    x^{'} &= x                      \\ y^{'} &= (\delta \times y)/\sqrt{(R^2-x^2)}\\ \label{eq3}
    z^{'} &= z.
\end{align}

Computing the Jacobian transfer matrix, $J$, for the above coordinate transformation, we can compute the primed material:
\begin{equation}\label{equ:erprime}
   \epsilon_r^{'} = \frac{J\epsilon_rJ^{'}}{\det{J}},
\end{equation}
\noindent which can be presented in terms of the original coordinates as shown in (\ref{eq:ERxx}) and (\ref{eq:ERyy}):

\begin{center}
\begin{align}
    \epsilon_{rxx}^{'} &= \epsilon_{rzz}^{'} = -\sqrt{R^2 - x^2} \frac{\frac{x^2 + y^2}{R^2} - 2}{\delta} \label{eq:ERxx}\\ 
    \epsilon_{ryy}^{'} &\triangleq 1 \label{eq:ERyy}
\end{align}
\end{center}
The resulting permittivity map is shown in Fig.\,\ref{fig:flatlens}, from which we can see the length of the lens remains the same while the width is squeezed to $\delta$. The permittivity distribution looks similar but is compressed. The electrical thickness must be much greater in the squeezed direction, which results in a maximum permittivity of $2/\delta$. In our design shown in Fig.\,\ref{fig:flatlens}, the maximum permittivity corresponds to $\epsilon_{rmax}=12$ at the center with a squeeze ratio of $1/6$. The minimum permittivity is still $\epsilon_{rmin}=1$ at the edges.

\begin{figure}
    \centering
    \subfloat[Hex]{
        \label{unitcell:hexhex}
        \includegraphics[width=0.225\columnwidth]{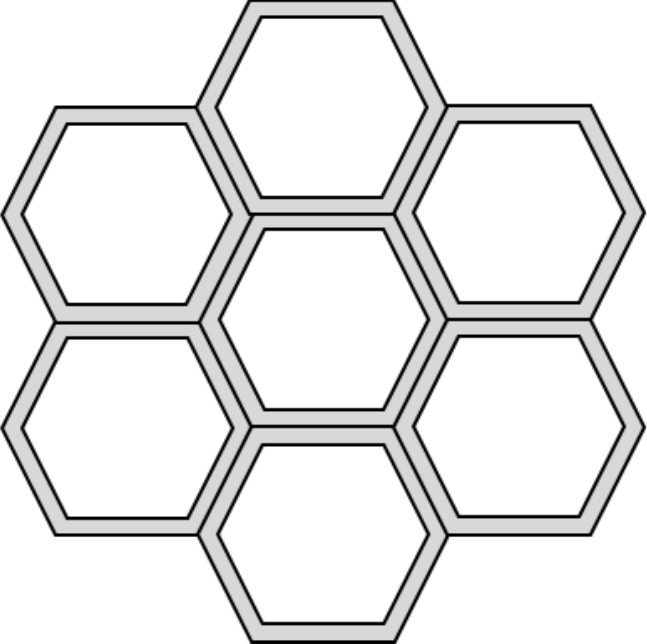}
        }
    \subfloat[Tri]{
        \label{unitcell:circletriangle}
        \includegraphics[width=0.25\columnwidth]{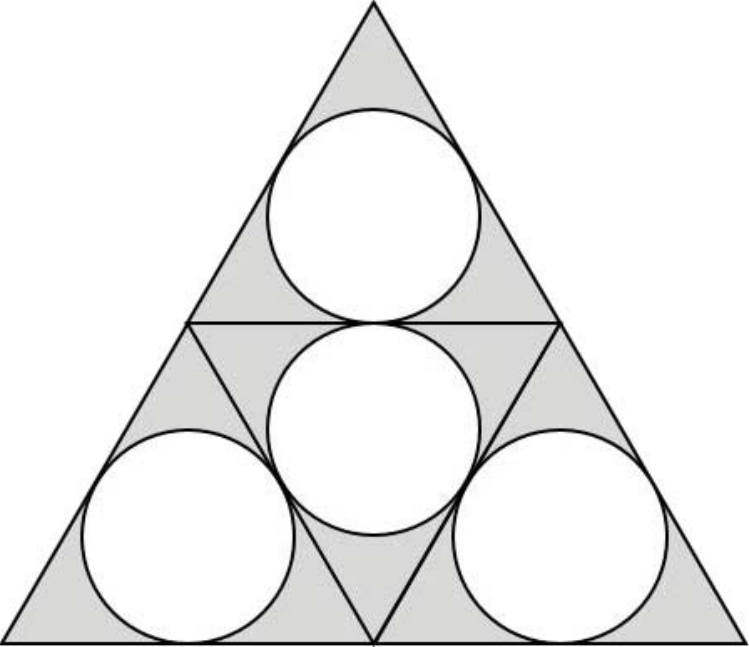}
        }
    \subfloat[Squ]{
        \label{unitcell:circlesquare}
        \includegraphics[width=0.2\columnwidth]{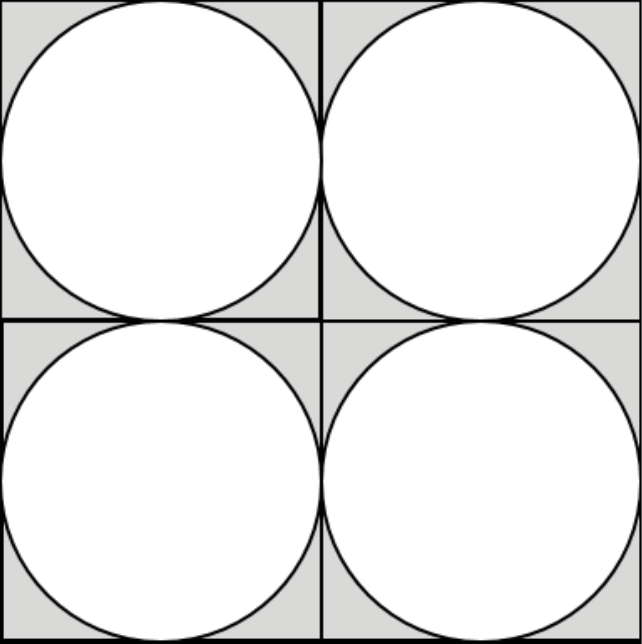}
    }
    \caption{Various perforated dielectric unit cells are compared. (a) A matched $n$-gon perforation and lattice (here a hexagonal void and lattice). (b) A circular perforation on a triangular lattice. (c) A circular perforation on a square lattice.}
    \label{unitcell}
\end{figure}

The lens design is now fully prescribed in Fig\,\ref{fig:flatlens} and we turn to the task of fabrication. By realizing the permittivity gradient with a perforated silicon dielectric, an effective permittivity can ideally range from 1 to 11.8. In the original work on perforated dielectrics \cite{potosa_perforateddielectric_1994} the medium was mechanically drilled to make circular voids on a square lattice (indicated with ``Squ'', see Fig.\,\ref{unitcell:circlesquare}) and circular voids on a triangular lattice (indicated with ``Tri'', see Fig.\,\ref{unitcell:circletriangle}) in a background dielectric. The effective permittivity of a perforated dielectric can be approximated as
\begin{equation}\label{eq:ereffective}
    \epsilon_{\textrm{\tiny eff}} = \epsilon_r\left(1-\alpha\right)+\alpha,
\end{equation}
\noindent where $\epsilon_r$ is the relative permittivity of the background dielectric and $\alpha$ is the filling factor equal to the ratio of the void area to the unit-cell area. For circular unit-cells on square and triangle lattices the minimum fill factor is equal to $\frac{\pi}{4}$ and $\frac{\pi}{2\sqrt{3}}$, respectively. Recall that the permittivity map contains unit-cells with $\epsilon_{r}=1.0$. We will show, later, that achieving a minimum permittivity close to $1.0$ is crucial for wide angle beam-steering of the lens. If the background permittivity is that of silicon, $\epsilon_r=11.8$, the corresponding minimum effective permittivity of the ``Squ'' and ``Tri'' configurations are 3.3 ($28\%$ of the background permittivity) and 2.0 ($17\%$ of the background permittivity) respectively, which significantly limits the fabrication of many TO designs. 

\begin{figure}[t]
\begin{center}
\noindent
  \includegraphics[width=0.8\columnwidth]{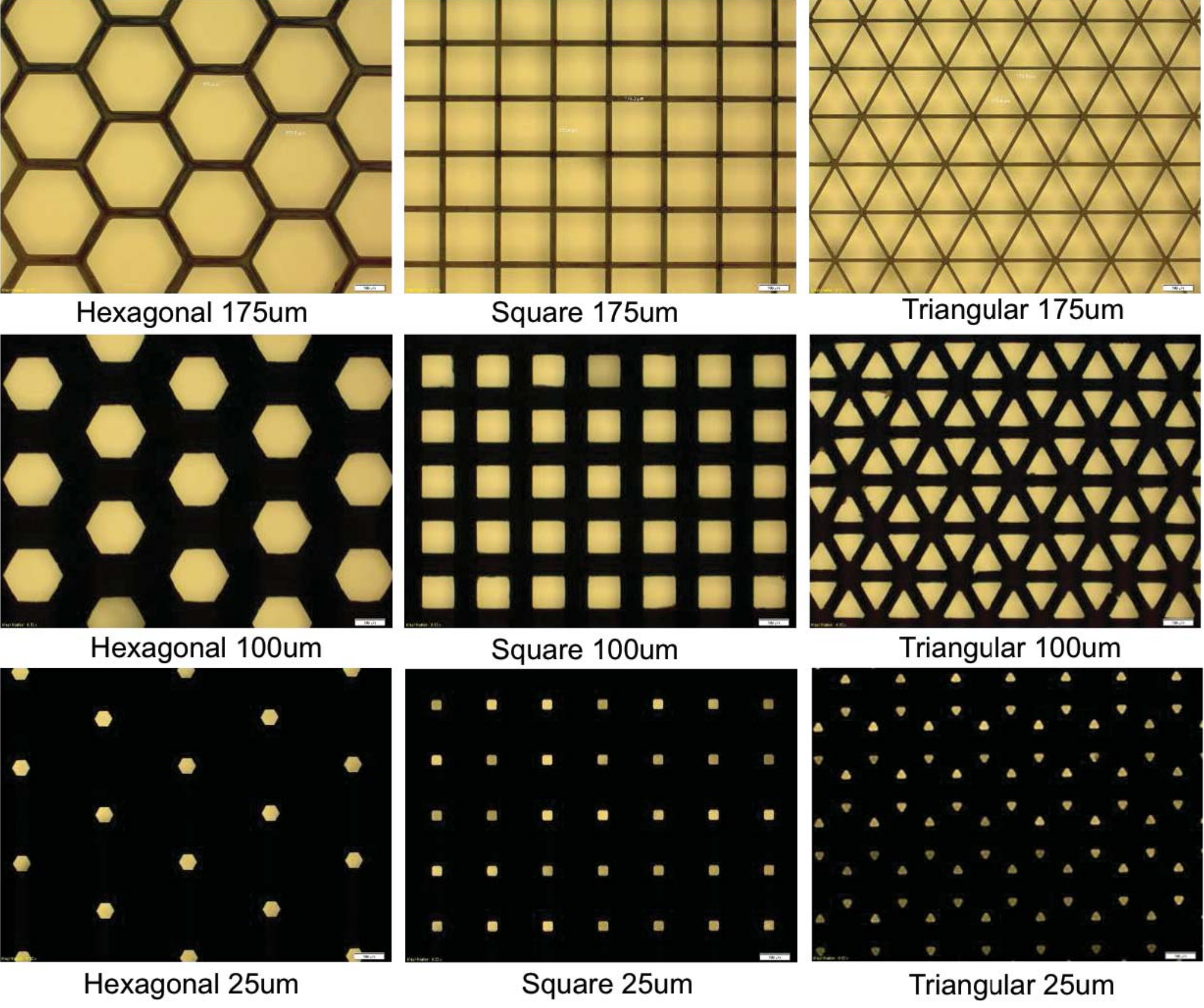}
  \caption{From \cite{garcia_apsursi}: Hexagonal, square, and triangular perforations with characteristic dimension $175\mu$m, $100\mu$m, and $25\mu$m as seen with backlight at 10x magnification. All perforations are on a regular $200\mu$m lattice in bulk silicon.}
  \label{fig:Fabrication}
\end{center}
\end{figure}

\begin{figure}[tb]
\begin{center}
  \includegraphics[width=0.9\columnwidth]{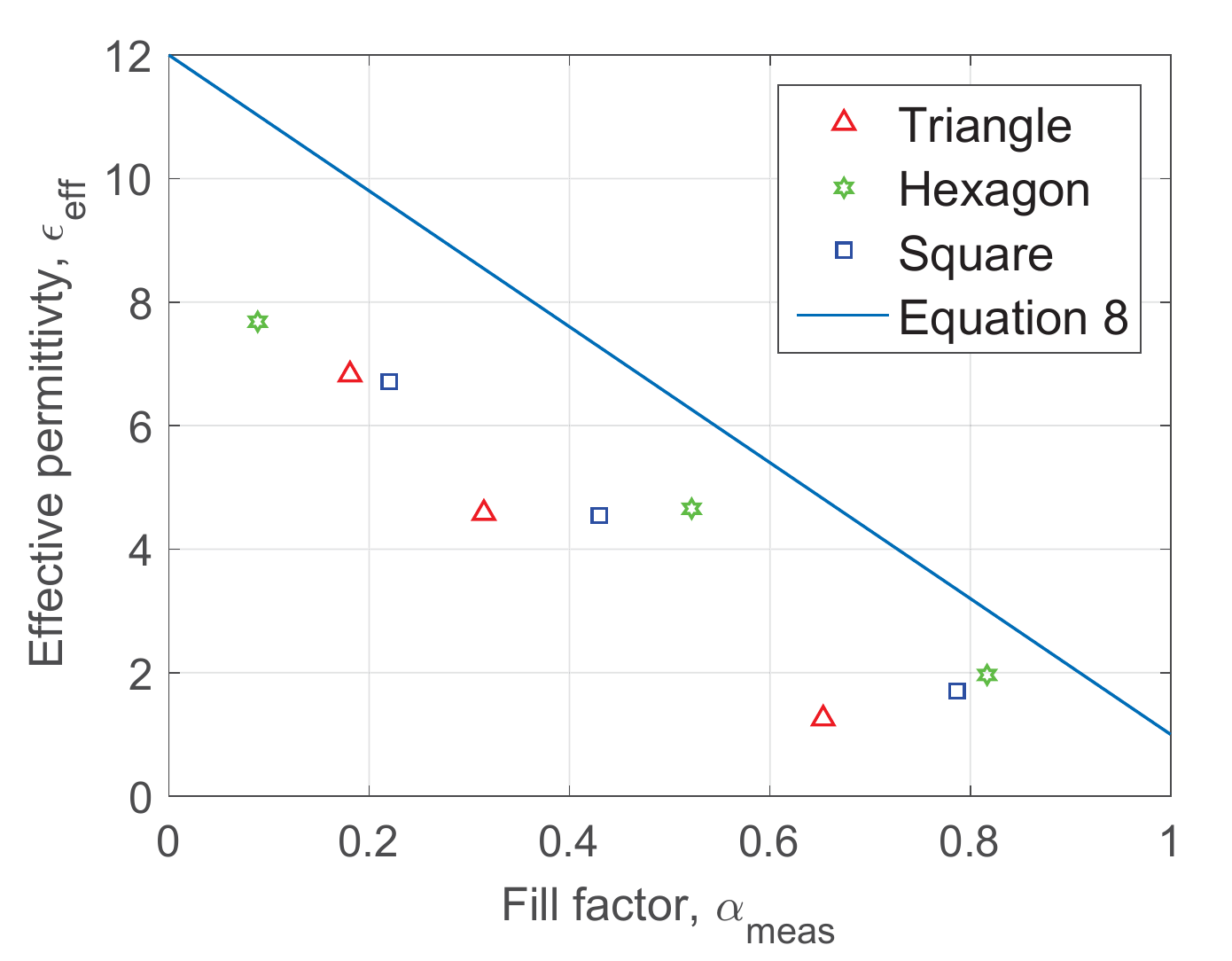}
  \caption{From \cite{garcia_apsursi}: Average permittivity of each geometry at each feature size.}
  \label{fig:Results}
\end{center}
\end{figure}

In order to approach $\epsilon_{rmin}=1.0$ we have proposed the use of $n-$gon voids on matched lattices and demonstrated their feasibility \cite{garcia_apsursi}. Figure\,\ref{unitcell:hexhex} shows a hexagonal void on a hexagonal lattice. Theoretically the minimum permittivity can approach $1.0$ as the volume of unetched dielectric becomes asymptotically small. We have fabricated silicon wafers with permittivities as low as $\epsilon_r=1.25$, or $10\%$ of the background permittivity. Figure\,\ref{fig:Fabrication} shows square, triangular, and hexagonal unit-cells on matched lattices. These structures were examined for robustness to manufacture and for their ability to exhibit a minimum permittivity. Features are exposed on an undoped silicon wafer with a diameter of $25.4$mm and a thickness of $270\mu$m. Etching is performed with a Bosch process. The resulting permittivities are shown in Fig. \,\ref{fig:Results} where the minimum permittivity value for the triangle structure is 1.25 which agrees with the the permittivity map constraints discussed above. Current Bosch etches exhibit significant undercut; fixing this problem will improve etch control.

\section{Wide-angle Beam-steering of Realistic Lenses}

In this section we use Ansys HFSS to explore the performance limitations of such a beam-steering lens antenna under the above fabrication limitations. Our objective is to determine the importance of permittivity range and in particular minimum effective permittivity to achieve wide angle beam-steering. 

If we constrain our fabrication to the three unit-cell structures shown in Fig.\,\ref{unitcell} we will be limited to minimum permittivities of $28\%$, $17\%$, and $10\%$ of the background permittivity for the ``Squ'', ``Tri'', and ``Hex'' configurations respectively. For a silicon background of $\epsilon_r=11.8$, the minimum permittivities are $3.3$, $2.0$, and $1.25$. Now truncating the permittivity map from Fig.\,\ref{fig:flatlens} with these values we can count the percentage of unit-cells that are effected by the fabrication limitations as summarized in Table \ref{table:1}. The background $\epsilon_r$ corresponds to base lens material of silicon ($\epsilon_r=11.8$), and two potentially higher permittivities of $\epsilon_r=20$ and $\epsilon_r=50$ (corresponding to lens compression ratios of $~6$, $10$ and $25$, respectively). Table \ref{table:1} shows that for a silicon lens on a square lattice where the minimum fabricated permittivity is $\epsilon_{rmin}=0.28\epsilon_r=3.3$, $~12\%$ of the unit cells in the idea design will have required permittivities below that of the fabrication limit. However, the same lens fabricated on a hexagonal lattice can realize a minimum fabricated permittivity of $\epsilon_{rmin}=0.10\epsilon_r=1.25$, resulting in only $2\%$ of prescribed cells being below the minimum fabrication limit. Lenses fabricated with a fewer number of prescribed cells below the fabrication limit will more faithfully realize the desired lens response. Table\,\ref{table:1} shows that unit-cell structure while not thickness is the more important consideration for realizing the designed permittivity map.

\begin{table}
\begin{center}
\begin{tabular} { | m{1.6cm} | m{2cm} | m{1.6cm} | m{1.6cm} |} 
\hline
Background $\epsilon_r$ & $\textrm{Square lattice}$ ($\epsilon_{rmin}=0.28\epsilon_r$) & Triangle lattice (0.17$\epsilon_r$) & Hexagon lattice (0.10$\epsilon_r$)\\ 
\hline
$\epsilon_r = 11.8$(Si) & 12.29 & 5 & 2.02 \\ 
\hline
$\epsilon_r = 20$ & 12.7 & 5.164 & 1.844 \\ 
\hline
$\epsilon_r = 50$ & 12.67 & 5.16 & 1.84 \\
\hline 
\end{tabular} 
\vspace{0.1cm}
\caption{Percentage of unit cells lower than the fabrication limit for different lattice structures}
\label{table:1}
\end{center}
\end{table}

\begin{figure}[h!]
    \centering
    \includegraphics[width=0.85\columnwidth]{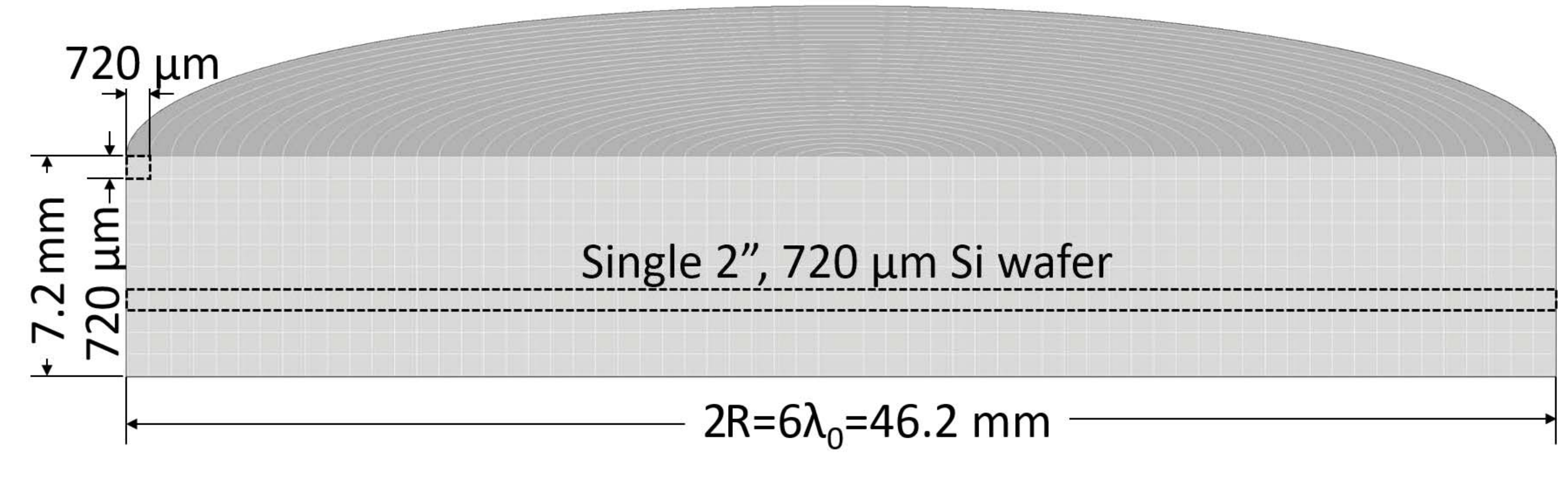}
    \caption{Model for electromagnetic simulations.} \label{fig:hfssmodel}
\end{figure}

Figure\,\ref{fig:hfssmodel} shows half of the final lens used for full-wave electromagnetic simulations. It is a rotational extrusion of the truncated permittivity maps. The lens is comprised of a $64 \times 10$ array of unit-cells of size $720\mu m \times 720\mu m$. The lens diameter is equal to $6\lambda$ where $\lambda$ is $7.7$mm, the free-space wavelength at $39GHz$. The associated thickness is $0.94\lambda=7.2$mm. The focal point is $7.7$mm which is almost identical to that of the typical Luneburg lens. 

\begin{figure}[h!]
    \centering
    \includegraphics[width=0.85\columnwidth]{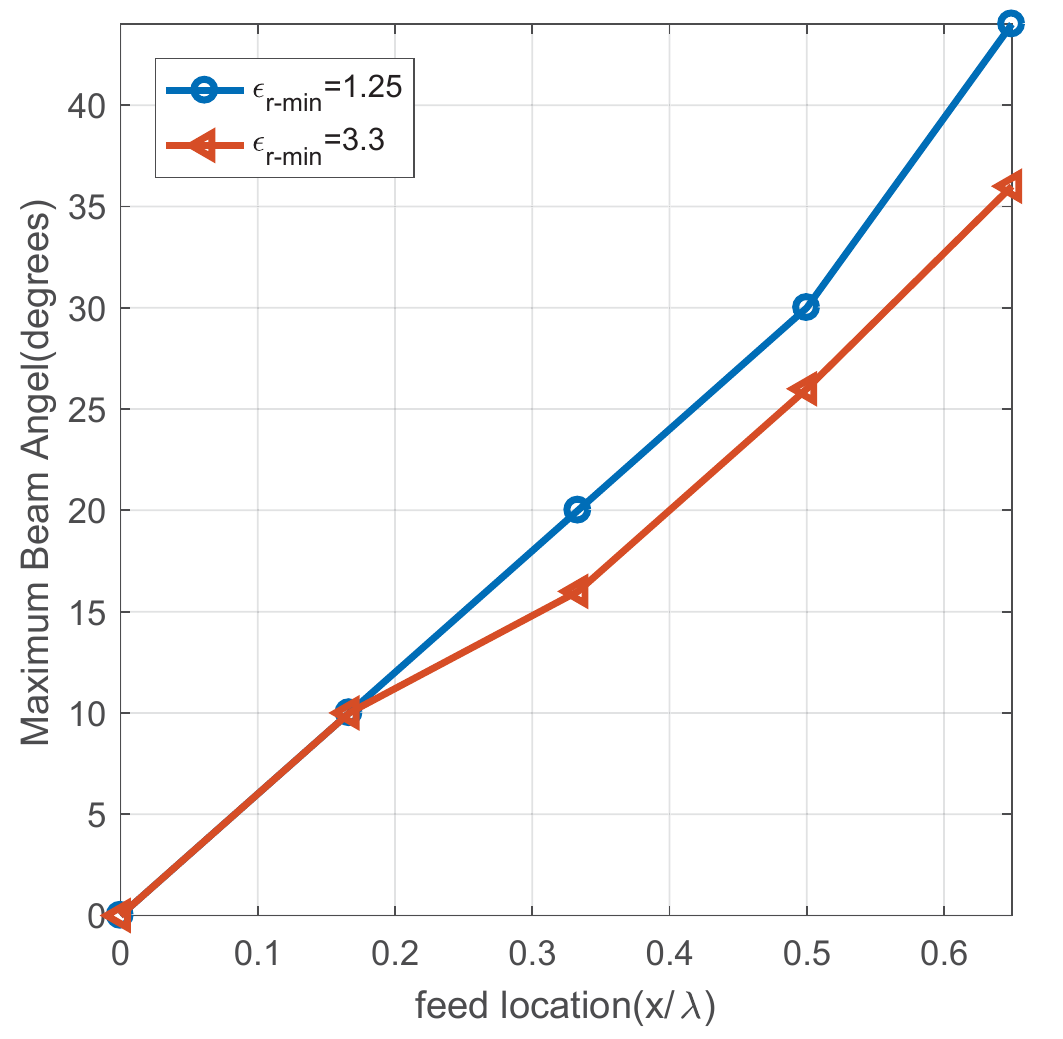}
    \caption{Maximum steered angle versus feed location for lenses with minimum $\epsilon_r$ of 1.25 and 3.3, respectively}
    \label{fig:anglevsfeed}
\end{figure}

Figure\,\ref{fig:anglevsfeed} shows the maximum beam steering angle versus feed location for the silicon lens fabricated with ``Squ'' and ``Hex'' unit-cell configurations with minimum permittivity of $3.3$ and $1.25$, respectively. At shallow scan angles the two lenses perform identically because the majority of the energy propagates throughout the center of the lens where permittivity is much higher than $\epsilon_{rmin}$. However, as beam scan angle increases the ``Hex'' unit-cell with $\epsilon_{rmin}=1.25$ outperforms the ``Squ'' unit-cell. This is because as scan angle increases the edge of the lens plays a more prominent role in beam-steering and the edge of the lens contains the lower permittivity unit-cells. 


\section{Conclusion}

In this paper, we present a way to design lens antennas by transformation optics. We have shown that by using an $n-$gon void on a matched lattice we can achieve a minimum permittivity of $1.25$ which yields a maximum beam-steering angle of $44$ degrees. This technology is a promising candidate for enabling practical low-loss, low-cost and compact beam-steering lens antennas for mmWave MIMO with wide beam-steering angles.
\bibliographystyle{IEEEtran}

\bibliography{vtc_2017.bbl}

\end{document}